\documentclass[a4paper,fleqn,usenatbib]{mnras}

 \usepackage{lineno}
\usepackage{newtxtext,newtxmath}

\usepackage[T1]{fontenc}
\usepackage{ae,aecompl}

\usepackage{graphicx}	
\usepackage{amsmath}	
\usepackage{amssymb}	

\title[CME impact on comet 67P]{CME impact on comet 67P/Churyumov-Gerasimenko}

\author[N. J. T. Edberg et al.]{
Niklas J. T. Edberg,$^{1}$\thanks{E-mail: niklas.edberg@irfu.se (NJTE)}
M. Alho,$^{2}$
M. Andr\'e,$^{1}$
D. J. Andrews,$^{1}$
E. Behar,$^{3}$
J. L. Burch,$^{4}$\newauthor
C. M. Carr,$^{5}$
E. Cupido,$^{5}$ 
I. A. D. Engelhardt,$^{1}$
A. I. Eriksson,$^{1}$
K.-H. Glassmeier,$^{6}$\newauthor
C. Goetz,$^{6}$
R. Goldstein,$^{4}$
P. Henri,$^{7}$
F. L. Johansson,$^{1}$
C. Koenders,$^{6}$
K. Mandt,$^{4,8}$\newauthor
H. Nilsson,$^{3}$
E. Odelstad,$^{1}$
I. Richter,$^{6}$
C. Simon Wedlund,$^{9}$
G. Stenberg Wieser,$^{3}$\newauthor
K. Szego,$^{10}$
E. Vigren,$^{1}$
and M. Volwerk$^{11}$
\\
$^{1}$Swedish Institute of Space Physics, Uppsala, Box 537, SE-75121, Sweden\\
$^{2}$Aalto University, School of Electrical Engineering, Department of Radio Science and Engineering, PO Box 13000, 00076 Aalto Finland\\
$^{3}$Swedish Institute of Space Physics, Box 812, 981 28 Kiruna, Sweden\\
$^{4}$Southwest Research Institute, 6220 Culebra Rd., San Antonio, TX 78238, USA\\
$^{5}$Imperial College London, Exhibition Road, London SW7 2AZ, United Kingdom\\
$^{6}$TU - Braunschweig, Institute for Geophysics and extraterrestrial Physics, Mendelssohnstr. 3\\
$^{7}$Laboratoire de Physique et Chimie de l'Environnement et de l'Espace, Orl\'{e}ans Cedex 2, France\\
$^{8}$Department of Physics and Astronomy, University of Texas at San Antonio, San Antonio, TX, USA\\
$^{9}$Department of Physics, University of Oslo, Box 1048 Blindern, 0316 Oslo, Norway\\
$^{10}$Wigner Research Center for Physics, Budapest, Hungary\\
$^{11}$Space Research Institute, Austrian Academy of Sciences, Schmiedlstrasse 6, 8042 Graz, Austria\\
}

\date{}

\pubyear{}

\begin{document}
\label{firstpage}
\pagerange{\pageref{firstpage}--\pageref{lastpage}}
\maketitle

\begin{abstract}
We present Rosetta observations from comet 67P/Churyumov-Gerasimenko during the impact of a coronal mass ejection (CME). The CME impacted on 5-6 Oct 2015, when Rosetta was about 800 km from the comet nucleus, \textcolor{black}{and 1.4 AU from the Sun}. Upon impact, the plasma environment is compressed to the level that solar wind ions, not seen a few days earlier when at 1500 km, now reach Rosetta. In response to the compression, the flux of suprathermal electrons increases by a factor of 5-10 and the background magnetic field strength increases by a factor of $\sim$2.5. The plasma density increases by a factor of 10 and reaches 600 cm$^{-3}$, due to increased particle impact ionisation, charge exchange and the adiabatic compression of the plasma environment. We also observe unprecedentedly large magnetic field spikes at 800 km, reaching above 200 nT, which are interpreted as magnetic flux ropes. We suggest that these could possibly be formed by magnetic reconnection processes in the coma as the magnetic field across the CME changes polarity, or as a consequence of strong shears causing Kelvin-Helmholtz instabilities in the plasma flow. Due to the \textcolor{black}{limited orbit of Rosetta}, we are not able to observe if a tail disconnection occurs during the CME impact, which could be expected based on previous remote observations of other CME-comet interactions.
\end{abstract}




\section{Introduction}
The Rosetta spacecraft arrived at comet 67P/Churyumov-Gerasimenko in Aug 2014, when at 3.6 AU from the Sun. Since then, it followed the comet at a distance of 10-1500 km from the nucleus while the comet passed through perihelion, at 1.2 AU, and outward again until end of mission \textcolor{black}{in Sep 2016 when at 3.8 AU from the Sun}. As a comet moves closer to the Sun it heats up and the outgassing rate increases. The neutrals together with the ionized particles and the dust lifting from the surface of the nucleus, builds up a cometary coma. The structure and dynamics of the plasma environment of the coma of 67P have been explored extensively since arrival, using measurements from the Rosetta Plasma Consortium (RPC) instrument suite \citep{carr2007, glassmeier2007}. The bulk plasma in the coma is created mainly through photoionisation of the local neutral gas \citep{edberg2015,odelstad2015,vigren2015,galand2016}, but impact ionisation and charge exchange processes also contribute \citep{burch2015,simon2016}. Newly ionised particles immediately feel the presence of the solar wind convective electric field and are picked-up by the flow and start to gyrate \citep{goldstein2015}. The first observations of the comet plasma environment were reported by \citet{nilsson2015}, as cometary ions were measured when at 3.6 AU from the Sun and at a distance of $\sim$100 km from the nucleus. The flux of cometary ions as well as the local plasma density around the nucleus were observed to increase gradually as the comet moved closer to the Sun \citep{nilsson2015b, odelstad2015}. As the coma grows larger, the interaction with the solar wind becomes more pronounced. To ensure the conservation of momentum the solar wind bulk flow is accelerated in the opposite direction from the newly created ions \citep{broiles2015, behar2016}. Eventually, plasma regions and boundaries begin to form, which has also been shown in global 3-D hybrid and MHD simulations \citep{koenders2013,koenders2015,rubin2014,huang2016}. 

Due to the trajectory of Rosetta being in the close vicinity of the nucleus, the full solar wind interaction region has not been sampled throughout the mission. The plasma boundary closest to nucleus is the diamagnetic cavity, which has been observed, although intermittently \citep{goetz2016}. The diamagnetic cavity builds up as the neutral-ion friction force in the outgassing material exceeds the magnetic pressure force from the outside \citep{cravens1987}.  Also, as the coma grew larger around Rosetta, measurements indicated a transition from an inner region to an outer region over time, where the boundary in between was interpreted to be the collisionopause \citep{mandt2016}. The collisionopause is the boundary where the ions become collisional and piles up, and its location is dependent on both the neutral outgassing rate and velocity as well as the collision cross section. The inner region, within the collisionopause, shows significant dynamics in the plasma environment. Order of magnitude density variations to the hot/cold plasma mixture occur on timescales of seconds to minutes \citep{eriksson2016}.

Besides the continuous growth and decay of the coma as the heliocentric distance decreases and increases, respectively, the plasma environment of the comet also exhibits large variations due to the changing solar wind. \citet{edberg2016} studied four cases of impacting corotating interaction regions on the comet from October to December 2014, as the comet activity grew stronger, and \citet{mckenna-lawlor2016} observed two CMEs arriving at 67P in September 2014, i.e. soon after Rosetta's arrival at the comet when the outgassing was relatively low. These all impacted when outgassing rate was about $10^{25}$-$10^{26}$ particles s$^{-1}$ \citep{hassig2015}. The CIR impacts caused a compression of the plasma environment present, which led to increased fluxes of suprathermal electrons, increased ionisation rate, increased plasma density as well as an increase in the magnetic field strength.

CME impacts on other comets have only been observed remotely. During such observations only large-scale changes in the comets' comae and tails could be observed due to the limited resolution \textcolor{black}{of the images}. \citet{jones2004} observed how a CME impacted on comet 153P/Ikeya-Zhang and could study how the comet tail appeared scalloped when the varying interplanetary magnetic field draped around the comet. \citet{vourlidas2007} observed how a CME impact caused a tail-disconnection event in comet 2P/Encke. This was later modelled by \citet{jia2009} and it was suggested that the sudden magnetic field rotations associated with the CME  caused magnetic reconnection to occur in the tail of the comet, which was then subsequently disconnected. 

Here we will present in situ measurement from Rosetta during a CME impact on comet 67P when close to perihelion, to study the CME's effects on the local cometary plasma environment.

\subsection{Instruments}
\label{sec:instruments} 
In this paper we have used data from all five sensors of the RPC \citep{carr2007}. These are the Langmuir probe instrument (LAP) \citep{eriksson2007}, the mutual impedence probe (MIP) \citep{trotignon2007}, the magnetometer (MAG) \citep{glassmeier2007b}, the ion and electron sensor (IES) \citep{burch2007}, and the ion composition analyzer (ICA) \citep{nilsson2007}. For a detailed description of each instrument we refer to the individual instrument papers or, for a condensed summary, to the Instrument section in the multi-instrument study by \citet{edberg2016}. In brief, here we will use electron density and spacecraft potential measurements from the LAP1 sweeps, normally at cadency of 96 s or 160 s. The negative of the spacecraft potential is proportional to the logarithm of the electron density \citep{odelstad2015} and in the interval covered here, if assuming a fixed electron temperature, gives a good measure of the density.  The electron density from MIP is derived from the plasma frequency emission line, obtained in both Short and Long Debye Length modes (SDL and LDL, respectively). The short time scale density variations in MIP, often large, have been filtered out using a 5-minutes median filter and discarding times when the number of MIP measurements are considered too small to be representative of the actual average density. We have also used the vector magnetic field measurements from MAG at a cadence of up to 20 Hz as well as electron spectrograms from IES and ion spectrograms from ICA, separated in cometary (heavy) and solar wind (light) ions species. 

\section{Observations}
\subsection{Dayside excursion and CME impact}
\label{sec:excursion} 
\begin{figure}
	\includegraphics[width=\columnwidth]{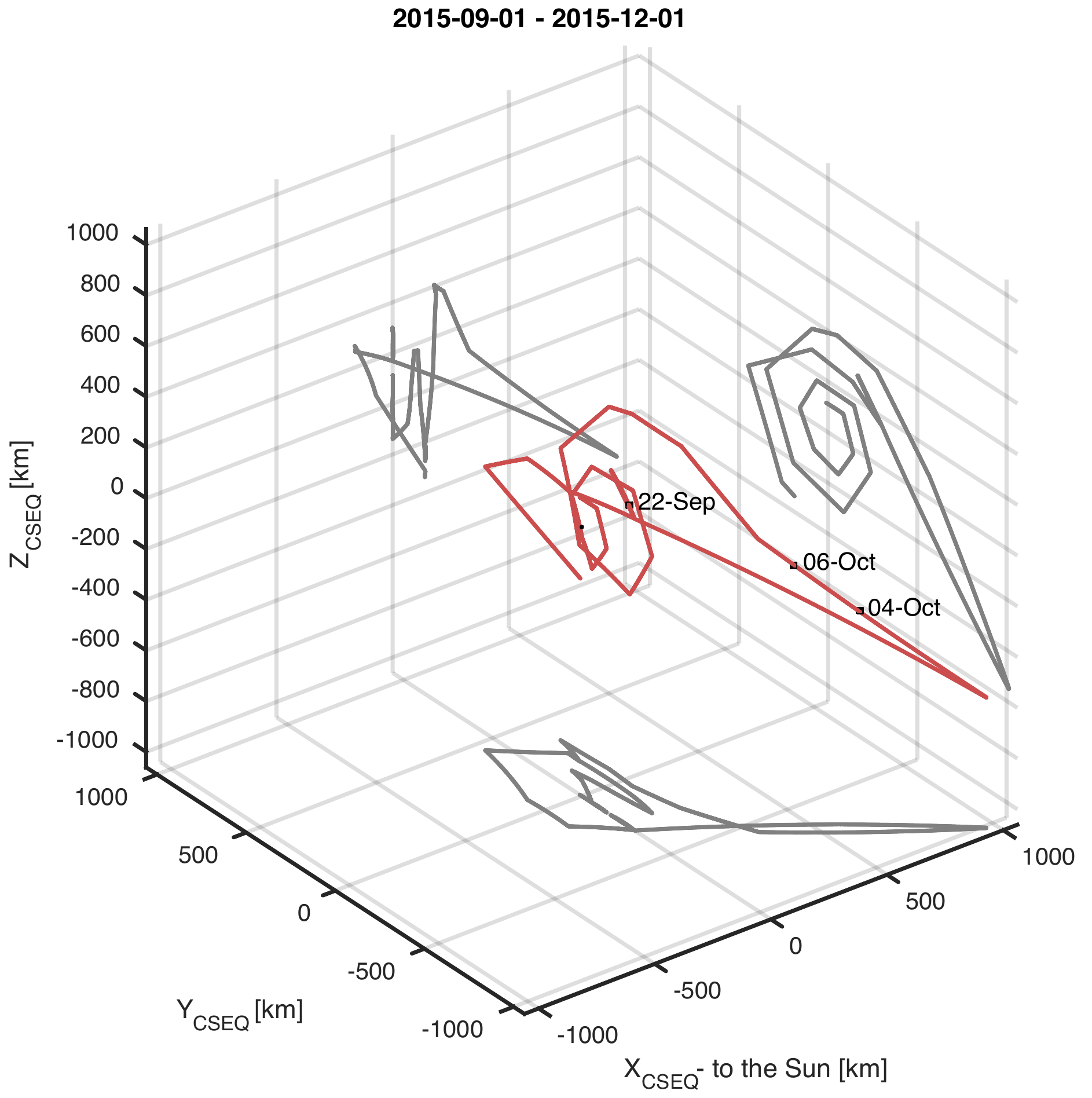}
    \caption{The trajectory of Rosetta in cometocentric CSEQ coordinates, with the projection on the three planes shown in grey. The interval includes the time of the Rosetta dayside excursion out to 1500 km. }
    \label{fig:exctraj}
\end{figure}

In September 2015, Rosetta left the near vicinity of the comet nucleus and began a two-week excursion outward in the coma to explore the spatial extent and structure of the plasma environment. During this interval the heliocentric distance spanned 1.34 - 1.41 AU. The trajectory is shown in Figure \ref{fig:exctraj} in the cometocentric solar equatorial coordinate system (CSEQ). In this system the x-axis points from the comet to the Sun, the z-axis is parallel to the component of the Sun's north pole orthogonal to the x-axis, and the y-axis completes the right-handed reference frame.
From being located in a near-terminator trajectory at around 300 km on 22 Sep 2015 Rosetta moved radially outward from the nucleus at an angle of about 50$^\circ$ to the comet-Sun line. Moving at a speed of $\sim$1 m/s relative to the comet Rosetta reached a distance of 1500 km on 30 Sep 2015 before slowly moving back in again. Meanwhile, SOHO images of the Sun captured \textcolor{black}{five individual coronal mass ejections (CME) on 30 Sep 2015, with the three largest ones being released around 06:00 UT, 09:36 UT and 10:00 UT. An image from SOHO can been seen in Figure \ref{fig:soho}.
\begin{figure}
	\includegraphics[width=\columnwidth]{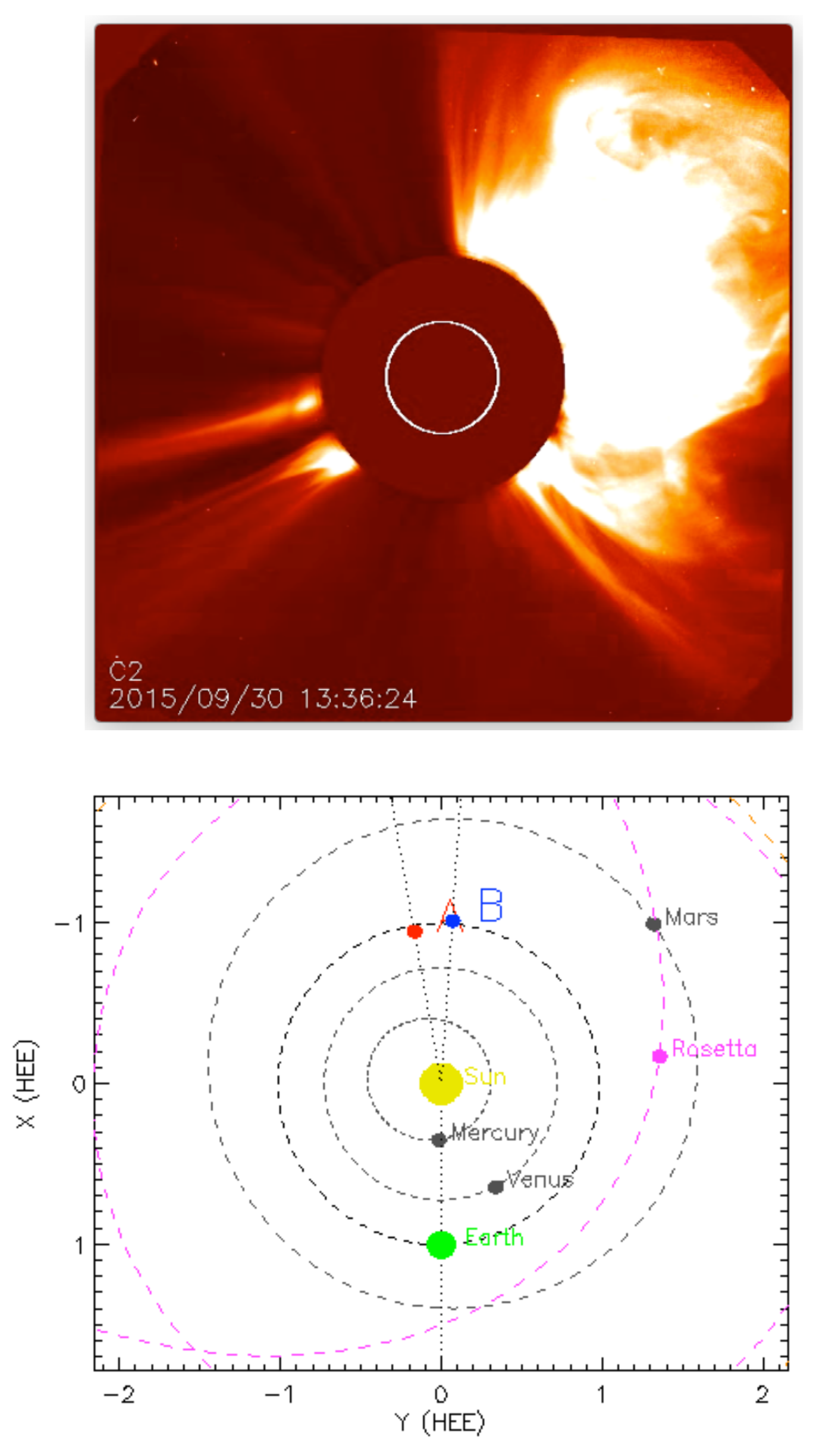}
    \caption{(Top) SOHO LASCO2 image of a CME being released on 30 Sep 2015 in the general direction of Rosetta (Image from \url{http://sohowww.nascom.nasa.gov/data/realtime/}). (Bottom) Position of the inner planets and Rosetta on 30 Sep 2015. A and B indicate the positions of the Stereo-A and Stereo-B spacecrafts, which did not observe the CME. There were unfortunately no spacecraft in operation at Mercury or Venus. At Mars, both the MAVEN and the MEX spacecraft were measuring the solar wind intermittently and saw a weaker signature of the CME (Figure from \url{http://stereo-ssc.nascom.nasa.gov/cgi-bin/make_where_gif}).}
    \label{fig:soho}
\end{figure}
The geometry of the Sun, Earth and Rosetta is shown in the lower panel of Figure \ref{fig:soho} and judging from this, the CMEs were released from the side facing comet 67P. From the SOHO image alone, showing only a projection of the CMEs as seen from the Earth's L1 Lagrange point, it cannot be determined exactly in which direction the CMEs were released. However, it is rather unlikely that the azimuthal extension of a CME is so narrow that it would not impact 67P in this configuration. The SOHO images indicate the angular widths of the projections of the three largest CMEs on this day to be above 82$^{\circ}$ (\url{http://cdaw.gsfc.nasa.gov/CME_list/UNIVERSAL/2015_09/univ2015_09.html}.} Mars was roughly at the same heliocentric distance at this time, but 30$^{\circ}$ off from the Sun-Rosetta line. Solar wind monitoring data at Mars indicate moderately disturbed solar wind signatures, while at Earth there was no indication of a CME arriving. At Venus and Mercury, no solar wind data was available at this time, unfortunately. \textcolor{black}{The three largest CMEs had linear velocities of 429 km/s, 586 km/s and 602 km/s, respectively, as determined from the SOHO images. It is not possible to accurately determine how the CMEs evolve with time and if they for instance merge with each other or slow down or accelerate \citep[see e.g.,][]{rollet2014}, which provides uncertainty in their evolution. If we assume that the three largest CMEs, released within 4 hours merge together to one single CME on its outward journey and assume a velocity of 500$\pm100$ km/s it would take roughly 4-6 days for it to reach Rosetta and 67P at 1.4 AU. It would then be arriving some time around 4-6 Oct 2015.} This coincides with the interval of the inbound leg of Rosetta's dayside excursion. As will be shown next, comprehensive evidence is found that the CME does indeed impact comet 67P, and significantly affects the plasma environment. 

\begin{figure}
	\includegraphics[width=19 cm]{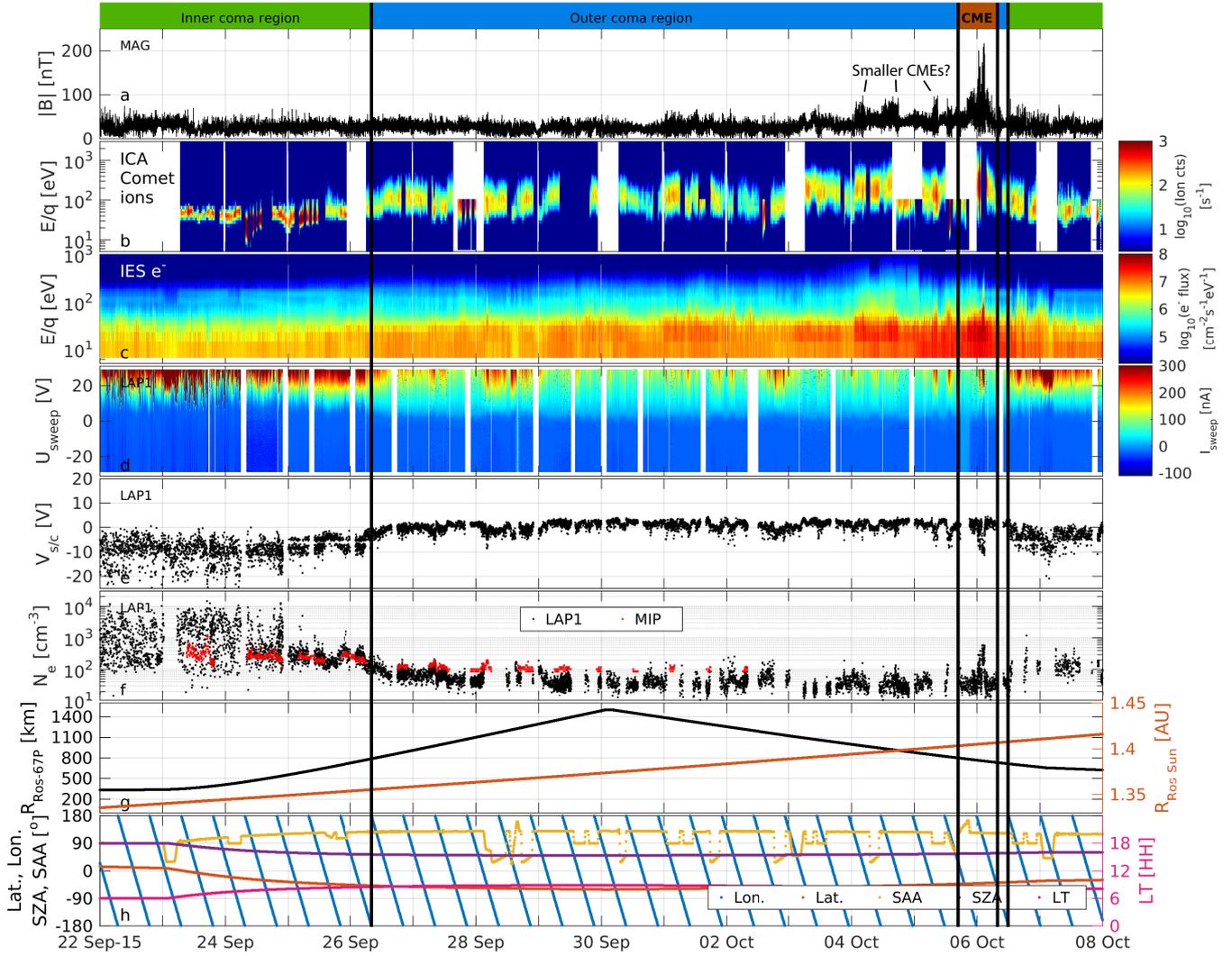}
    \caption{Overview of RPC data from the dayside excursion and especially including the interval of the CME impact. The panels show (a) the magnetic field strength, (b) spectrogram of ICA measured heavy (cometary) ions, (c) spectrogram of IES measured electrons (d) LAP sweeps (current collected is colour-coded, bias potential fed to the probe on the vertical axis) (e) spacecraft potential measured by LAP (f) electron density measured by LAP (black dots) where data points during spacecraft attitude changes are excluded together with MIP density estimates (red dots), (g) distance to the comet and the Sun from Rosetta (h) longitude and latitude (projected down on a sphere), SZA, SAA (attitude angle) and Local time (LT) of Rosetta.}
    \label{fig:overview}
\end{figure}

Figure \ref{fig:overview} shows an overview of the RPC data gathered during the entire dayside excursion, including the interval 4-6 Oct 2015 \textcolor{black}{when the CME was expected to arrive}. We will first describe the RPC measurements during the dayside excursion before moving on to the detailed observations of the CME impact on the comet. \textcolor{black}{The magnetic field showed large variations throughout the interval with a background field strength (panel a) of about 25 nT, i.e. considerably stronger than the average IMF of a few nT, indicating that Rosetta was always located in the coma of the comet and not in the solar wind. At the start of the excursion Rosetta was located in the "inner" region of the plasma environment. Here, the energetic ($\gtrsim 10$ eV) ion and electron fluxes (panel b and c), were relatively low and the solar wind was completely shielded at this time. Any undisturbed solar wind ions, mainly H$^+$ flowing at 400 km/s, would have an energy of about 1000 eV. In this figure only cometary ions, which are separated out by their heavier masses, are shown. The ions and electrons observed here are all of cometary origin and have been created through ionisation of neutral particles from the comet and have energies of about 100 eV. The particle instruments are capable of measuring species with energies down to about 10 eV, below which the Langmuir probe instrument takes over in measuring the cold plasma properties. A redder color at higher energies indicates that more particles have been accelerated to higher speeds. The spread in energy at a particular time corresponds to the temperature of the plasma, which varies throughout the interval. The inner region was also characterised by highly variable Langmuir probe sweeps, spacecraft potential and plasma density, indicating that the properties of the colder populations of electrons and ions ($\lesssim 10$ eV) change rapidly (panel d, e and f). Order-of-magnitude changes occur on time scales of seconds to minutes.} When Rosetta was moving further out the local plasma environment changed. \citet{mandt2016} studied the structure of the plasma environment during this excursion in more detail and identified at least one type of boundary, interpreted as an collisionopause, first crossed at a distance of about 600 km during the outbound leg. This boundary crossing was observed, outbound, as an increase in energy and flux of the cometary ions and electrons, the spacecraft potential increased to positive values, the cold plasma density dropped an order of magnitude to about 10-100 cm$^{-3}$ and there was a moderate magnetic field increase at the same time. While simulations of the plasma environment had indicated that the cometary bow shock would be crossed before reaching 1500 km \citep{koenders2013,rubin2014}. \textcolor{black}{\citet{huang2016} used an MHD simulation to show that the bow shock was closer to 10000 km under perihelion conditions and illumination-driven neutral outgassing. It turned out that the cometary plasma environment was more extended than what Rosetta reached during the dayside excursion. In fact, the RPC measurements showed no indication of the presence of a bow shock, nor that of solar wind ions, once 1500 km was reached.}

Instead, Rosetta remained in this outer region, the ion-pile up region, for several days and at least until 00:00 on 4 Oct 2015. At this time Rosetta was at a distance of 1000 km and yet another increase in ion and electron energy and flux was observed (as opposed to a decrease as would be expected if crossing the collisionopause inbound again). This was accompanied by an increase in the magnetic field strength from an average of about 20 nT to an average of about 40 nT as well as a sudden increase in electron fluxes for energies $<200$ eV by a factor 2-5. The cold plasma density and the spacecraft potential remained unchanged at this time. These signatures are possibly purely due to the dynamics in the coma itself, \textcolor{black}{or some of the earlier, weaker but faster CMEs released from the Sun on 30 Sep.} The suprathermal electron fluxes as well as the ions and the magnetic field strength show several larger enhancements and decreases in the following 48 hours. 

At about 20:15 UT on 5 Oct 2015 the main impact of the CME occurs. This agrees well in time with when we expect the CME to arrive and the impact is clearly identified in all RPC data sets as an increase in magnetic field strength, plasma density, ion and electron flux. Before moving on to the detailed observations during the main impact we note that after the CME impact event around midnight on 5-6 Oct 2015, Rosetta is briefly located in the undisturbed ion pile-up region (i.e. outside the collisionopause) again for a few hours. Around noon on 6 Oct 2015 the collisionopause is finally crossed inbound as Rosetta continues to slowly move back toward the comet nucleus. 
 \clearpage
 \clearpage
 \clearpage
\begin{figure}
	\includegraphics[width=18 cm]{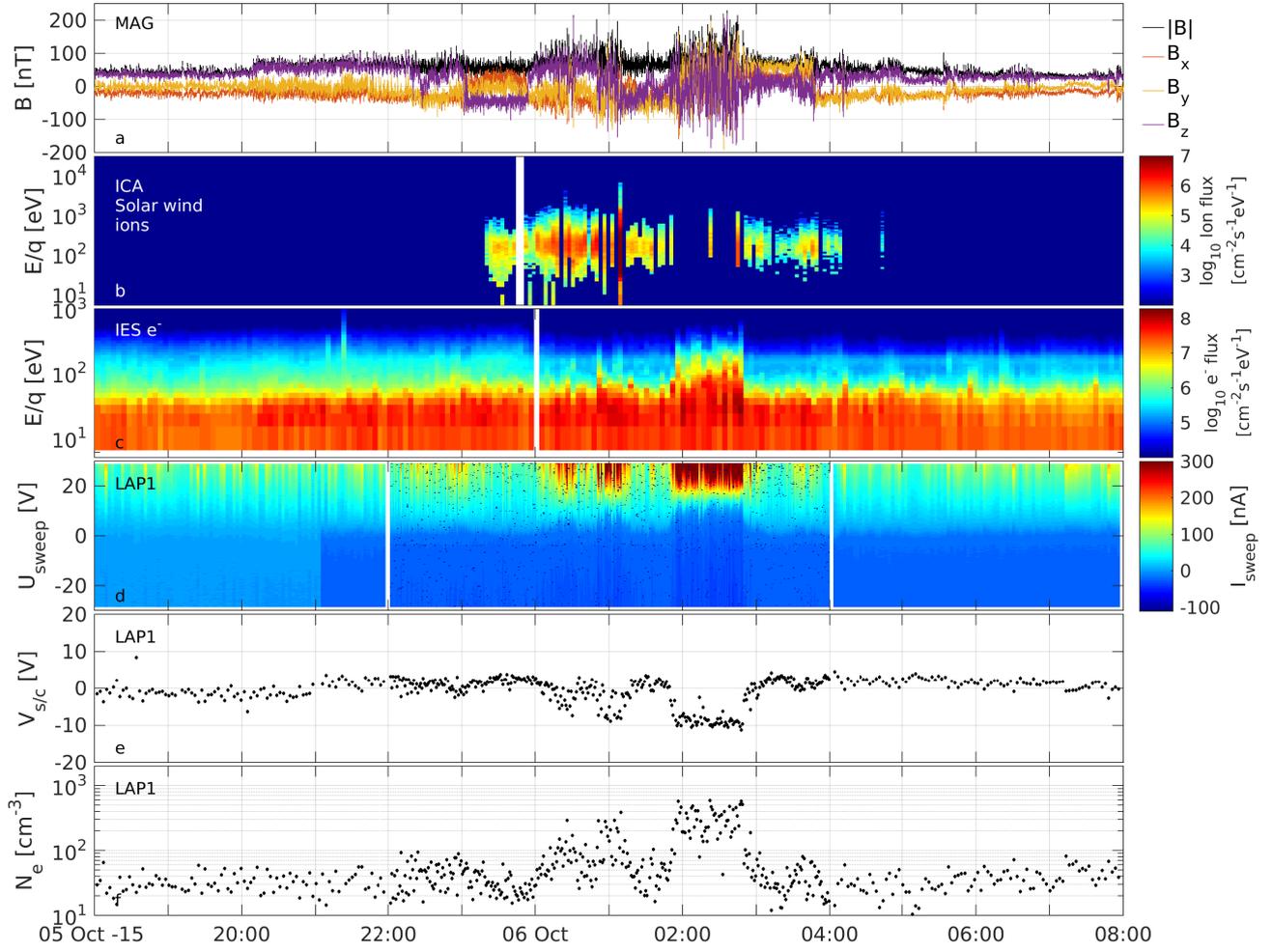}
    \caption{Time series of RPC data covering the interval of the CME impact. The panels show (a) magnetic field magnitude and components in CSEQ coordinates, (b) spectrogram of solar wind ions (c) spectrogram of electron fluxes (d) Langmuir probe sweeps, (e) spacecraft potential, and (f) electron density derived from the Langmuir probe sweeps. }
    \label{fig:zoom}
\end{figure}
\clearpage
 \subsection{CME influence on the comet}
Figure \ref{fig:zoom} gives a sub-set of the RPC data in order to study the detailed response to the CME impact on the comet. In the spectrogram in panel b we now show solar wind ions, rather than cometary ions as in the previous figure. \textcolor{black}{The ion instrument is capable of separating ions depending on their mass and charge and as the solar wind mainly consists of H$^+$ ions they are easily separated from the heavier cometary H$_2$O$^+$ and CO$^+_2$, CO$^+$ ions. As stated above the undisturbed solar wind would appear around 1000 eV, with some spread in energy, but here these H$^+$ ions appear at a much lower energy indicating that the solar wind has been slowed down by the cometary coma.} The magnetic field as well as the plasma data present several interesting features in this interval. Firstly, the CME appears to cause another increase in the magnetic field strength (panel a), from about 40 nT to about 60 nT at 20:10 UT and then to a maximum background field of about 100 nT around 02:00 on 6 Oct, i.e an increase of a factor of $\sim$2.5 from minimum to maximum during this interval. \textcolor{black}{The increases in magnetic field strength are accompanied by increases in the suprathermal electron flux as higher electron fluxes (redder color) are seen at energies from about 10-500 eV) (panel c). This is consistent with the CME causing a compression of the plasma environment. Furthermore, there are two 1-hour long intervals around 00:20-01:20 UT and 01:50-02:45 UT on 6 Oct 2015 when the magnetic field measurements show bursts of rapid fluctuations and high amplitude magnetic field spikes, some of which reach above 200 nT. }

During the two bursts, the LAP sweeps (panel d) and the derived plasma density and spacecraft potential (panel e and f) change significantly. The density increases by as much as a factor of 10 to reach 600 cm$^{-3}$, and the spacecraft potential drops from about +1 V to -10 V, indicating a significant increase in the flux of electrons. During the time of the CME impact, MIP was unfortunately operated in LDL mode, designed for plasma densities lower than about 300 cm-3, thus missing the CME itself. The large-scale magnetic field orientation changes, which occur over a time span of hours, are probably associated with the large-scale magnetic flux rope commonly seen across a CME. 

Furthermore, and of particular significance, the solar wind ions, which had been absent in Rosetta ICA and IES particle data since April 2015, were briefly observed again during this event, by the ICA instrument, from 23:00 UT on 5 Oct until 04:00 UT on 6 Oct (panel b). This suggests that the plasma environment had been compressed significantly, \textcolor{black}{such that the solar wind ions could briefly reach the detector}, and provides further evidence that these signatures in the cometary plasma environment are indeed caused by a solar wind event, such as a CME. 

Next, we will discuss particular effects of the CME impact in more detail, focusing on the solar wind ion observations in \ref{sec:solarwind}, the increased plasma density and fluxes in Section \ref{sec:density}, and finally the magnetic field spikes in Section \ref{sec:fluxrope}.

\section{Discussion}
\subsection{Penetration of the solar wind ions}
\label{sec:solarwind}
After April 2015, the solar wind ceased to be observed by Rosetta, which was located deep inside the coma \citep{mandt2016}. The solar wind did not reappear again until several months after perihelion. However, during the CME impact reported here the ICA instrument did in fact observe solar wind protons penetrating the coma. Panel b of Figure \ref{fig:zoom} shows a spectrogram of these solar wind ions. The solar wind ions (protons) have been slowed down during their path to Rosetta and end up at roughly the same energy as the cometary heavy ions (compare panels b in Figures \ref{fig:overview} and \ref{fig:zoom}). The similar energy unfortunately makes them harder to distinguish from each other by the ICA instrument. But as ICA is a mass-resolving instrument it is possible to separate the heavy cometary ions from solar wind ions. The solar wind ions are clearly observed here, but the fluxes are relatively low compared to e.g. solar wind spectra from before April 2016. In the interval 02:00-03:00 UT on 6 Oct there appears to be a gap in the solar wind observations. There is a significant decrease in the solar wind fluxes at this time but as the signal gets weaker the uncertainty in distinguishing them from cometary ions by the instrument also increases. The drop out of solar protons at this time is therefore to be regarded as a lower limit of the fluxes. The solar wind ions were observed to be deflected typically some $30^{\circ}-50^{\circ}$ from the comet-Sun line. The cometary ions have a preferred direction in the anti-sunward direction, but are scattered in their direction by a few tens of degrees. 

Rosetta was at this time at about 800 km from the nucleus. Earlier, when at 1500 km, i.e. at the furthest distance from the nucleus during the excursion, the solar wind was not seen at all. \textcolor{black}{The magnetic field strength also increased at the time of the CME impact, as discussed earlier, indicating a compression of the plasma in the coma. If interpreting the appearance of the solar wind ions as a pure compression, the CME then compressed the plasma environment to at least half its previous size on the dayside}. 

\subsection{Increased plasma density}
\label{sec:density}
As can be seen in panel f of Figure \ref{fig:zoom}, the cold plasma density (measured by LAP) increases by as much as one order of magnitude for two 1-hour long intervals in the morning of 6 Oct 2015. The spacecraft potential goes significantly negative, from +1V to -10V, indicating that the electron density must be increased, to provide a higher flux of electrons to the spacecraft. Alternatively, the electron temperature can be increased to provide the higher fluxes, but as LAP does not measure any increased temperature this does not appear to be the case. These signatures coincide with the large magnetic field spikes appearing and also with large-scale magnetic field rotations. At the same time, the flux of the more energetic electrons increases significantly, albeit more pronounced so for the second interval. The solar wind ions, on the other hand, do not appear to show a maximum of fluxes at the same time as the density peaks.

\begin{figure}
	\includegraphics[width=\columnwidth]{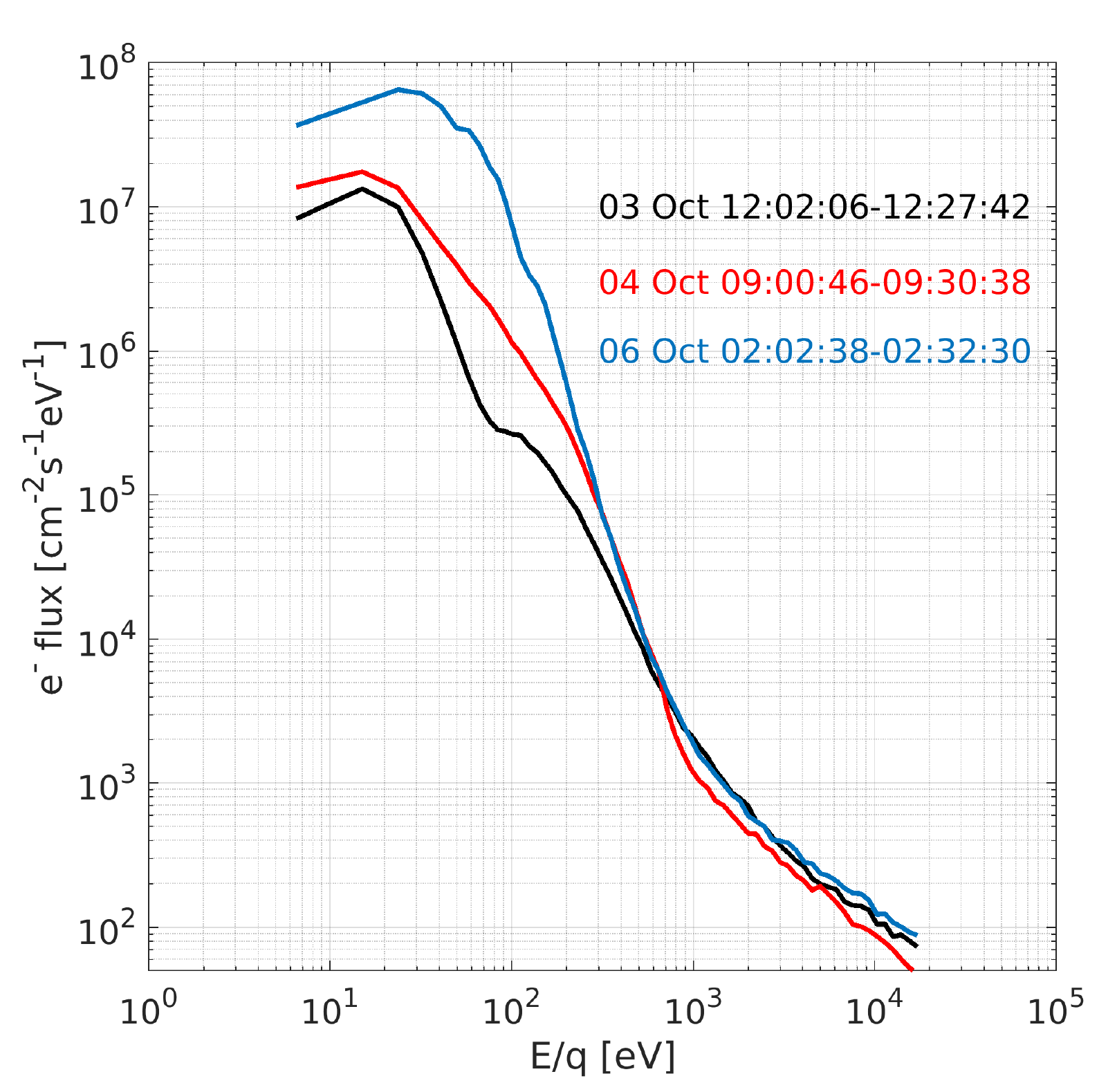}
    \caption{IES electron energy spectra during three intervals before and during the CME impact. Up to an order of 10 increase in electron fluxes for energies below 200 eV during the CME impact are observed during the CME impact (blue and red line), compared to before (black line).}
    \label{fig:3spec}
\end{figure}

To investigate if this density increase is due to increased impact ionisation from suprathermal electrons we calculate the ionisation frequency from three different suprathermal electron spectra in this time interval. Figure \ref{fig:3spec} shows the three spectra. The times are indicated in the figure and correspond to before CME impact, during elevated flux before the main CME impact, and during the time of maximum suprathermal electron fluxes. 
Combining these spectra with electron impact ionization cross sections of H$_2$O \citep{itikawa2005}, we obtain impact ionisation frequencies f$_{E}$ of $2.6\cdot10^{-8}$ s$^{-1}$,  $6.6\cdot10^{-8}$ s$^{-1}$, and  $6.8\cdot10^{-7}$ s$^{-1}$, respectively. For these calculations we have assumed isotropic electron fluxes and corrected the measured electron fluxes for the spacecraft potential as derived from LAP measurements. A more thorough treatment of electron impact ionisation for comet 67P can be found in \citet{galand2016}.
For comparison, the H$_2$O photoionization frequency is approximately $7\cdot10^{-7}/d^2 = 3.5\cdot10^{-7}$ s$^{-1}$, where $d$=1.41 AU is the heliocentric distance \citep{vigren2015}. If assuming proportionality between the total ionization frequency and the electron number density the enhanced electron impact ionization would only bring about a factor of~2.5 increase in the electron number densities. Hence, the observed increased density by a factor of ~10 can clearly not be attributed solely to increased particle impact ionization.

\begin{figure}
	\includegraphics[width=\columnwidth]{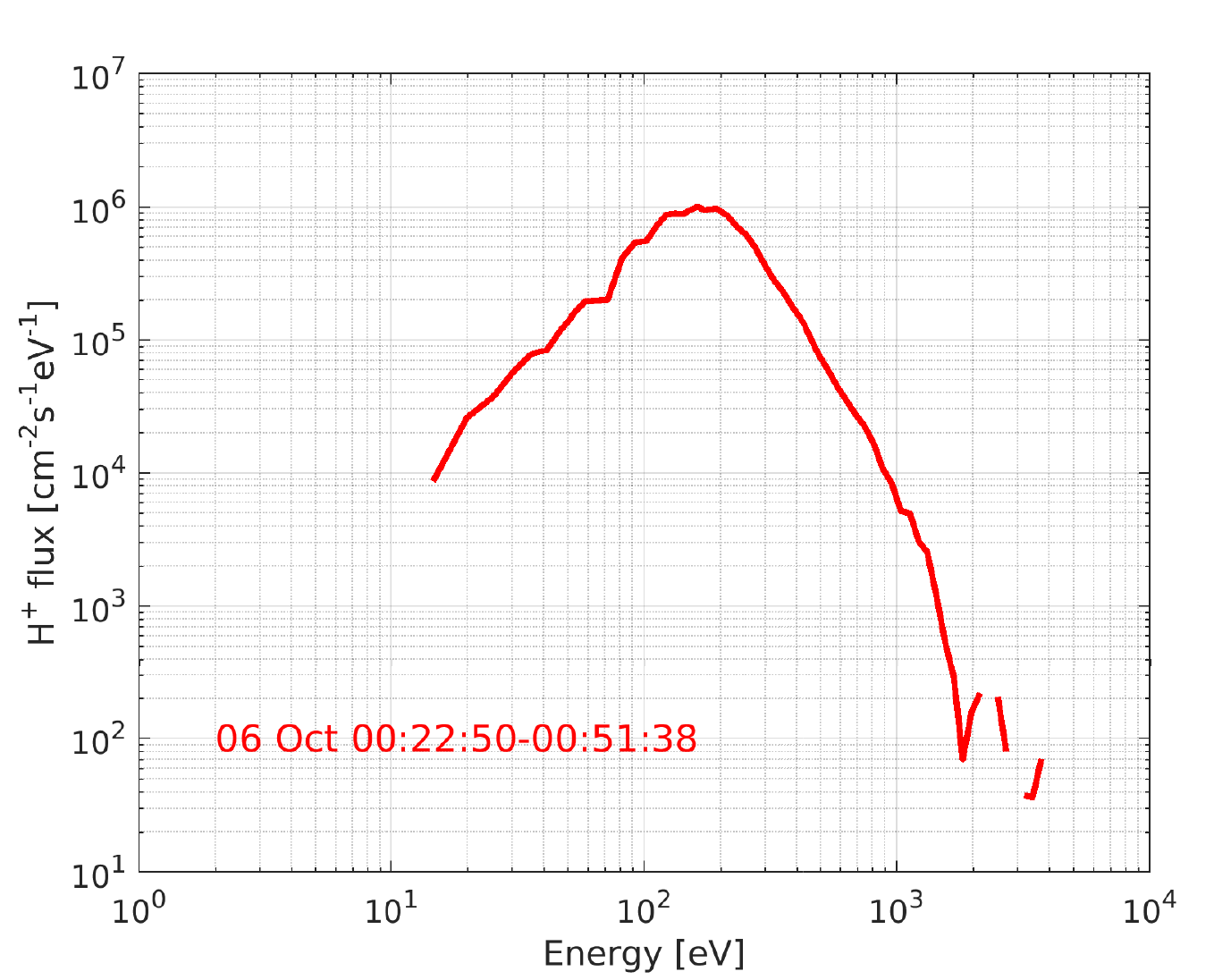}
    \caption{ICA solar wind ion energy spectra during the CME impact, when the LAP measured electron density was significantly increased. The solar wind is significantly reduced both in energy and flux as it reaches Rosetta.}
    \label{fig:icaspectra}
\end{figure}

Charge exchange is another process that could cause enhanced plasma density \citep{gombosi1987, burch2015}. Considering H$^+$ + H$_2$O $\rightarrow$ H + H$_2$O$^+$ to be the dominating charge exchange process \citep{simon2016}, we can estimate roughly its contribution to the increased density. An average energy spectra of solar wind ion flux (mainly protons) is shown in Figure \ref{fig:icaspectra}. This spectrum is averaged over an interval when the LAP measured electron density is increased to about 100 cm$^{-3}$. The  H$^+$ flux is typically F$_{H^+} \sim 1 \cdot 10^6$ cm$^{-2}$s$^{-1}$eV$^{-1}$ at maximum, at an energy $E_{max}$ of 200 eV. The charge exchange cross section $\sigma_{cx}$ is equal to $1.2\cdot10^{15}$ cm$^{-2}$ for H$^+$ with an energy of 200 eV \citep{mada2007}. The ion production rate (or charge transfer rate) is then f$_{cx} = \sigma_{cx} F_{H^+} E_{max} = 2.4 \cdot 10^{-7}$ s$^{-1}$, which is comparable to the photoionization and electron impact ionisation frequencies. However, charge exchange is not a net source of plasma, but rather changes the composition of some of the plasma from solar wind ions to heavier cometary ions. As the heavier ions will have lower velocity than the solar wind ions (in order to conserve momentum) there will be a pile up of plasma and an effective increase of the density. This increase depends on the local fraction of solar wind that is charge exchanging and it is challenging to calculate the exact density increase this would yield. More sophisticated models are needed, which is beyond the scope of this paper, and we leave that for any future study. We will settle with simply stating that there will most likely be a significant density increase caused by charge exchange at this time. However, we can also mention that preliminary results from hybrid models \citep{simon2016b} using variable input conditions to simulate the effects of a solar wind pressure pulse, such as the one studied here, are in tentative agreements with our results. Most importantly, in the simulation, the density may increase several times when the pressure pulse impacts \citep{alho2016}.
 
The compression due to the increased solar wind dynamic pressure cause another factor of 2.5 increase (\textcolor{black}{determined from the increase in background magnetic field strength, which we assume to be frozen into the plasma),} which brings us close to being able to explain the factor total of 10 increase in density, if taking into account the uncertainty of the simplified models in calculating the increased ionisation rates. We also note that the maximum electron impact ionisation (the blue spectra in Fig. \ref{fig:3spec}) does not occur when the measured solar wind flux are at maximum (blue spectra in Fig. \ref{fig:icaspectra}), such that the maximum effects of charge exchange and electron impact ionisation might not occur at the same time. It is also possible that some of the increased density is due to the changing field direction and that the cold plasma is accelerated by an electric field in the direction toward Rosetta, as discussed by \citet{vigren2015}. 

\subsection{Magnetic flux ropes in the coma}
\label{sec:fluxrope}
\begin{figure}
	\includegraphics[width=\columnwidth]{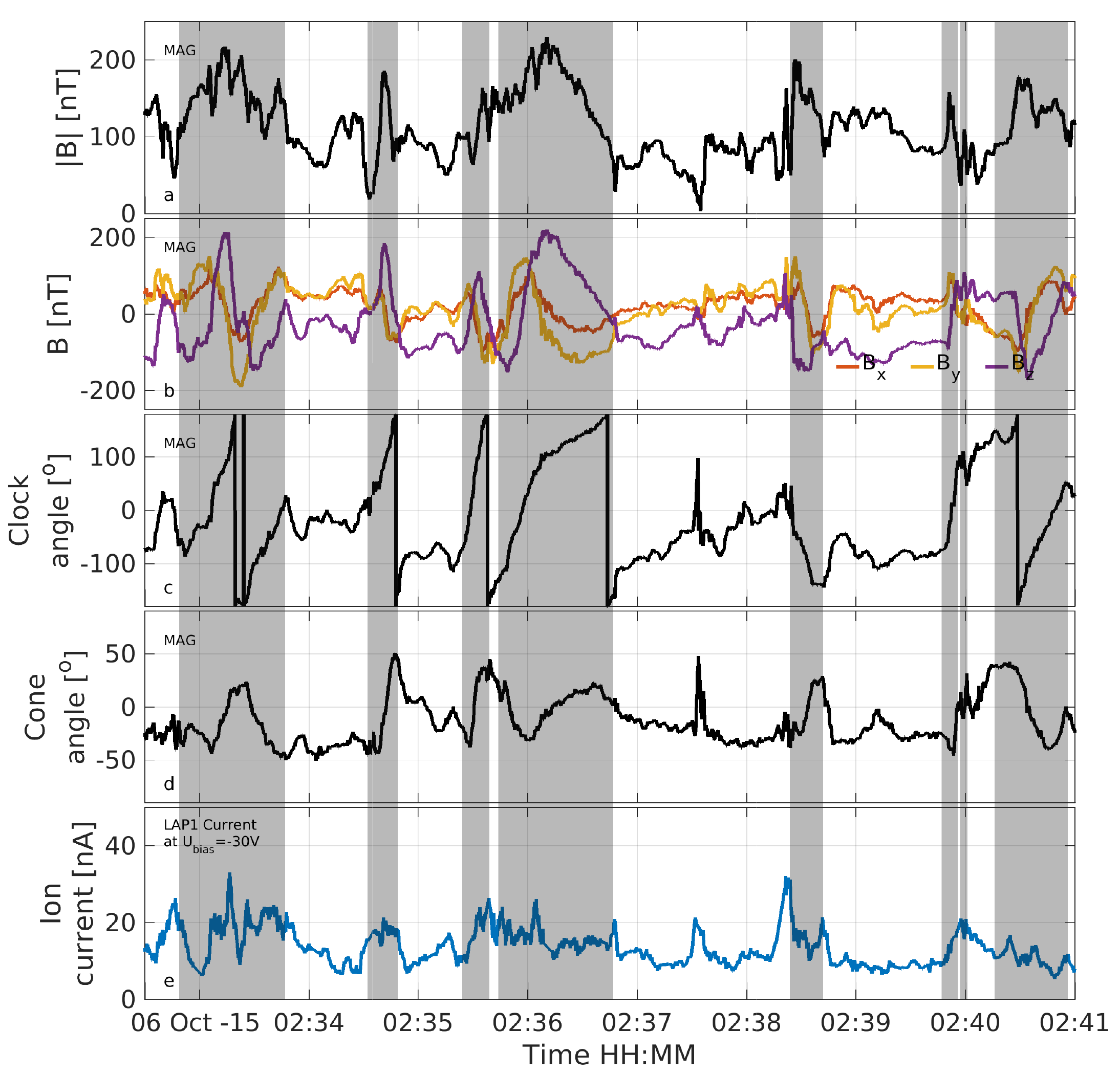}
    \caption{Time series of MAG (a) magnetic field magnitude, (b) components, (c) clock angle and (d) cone angle as well as (e) LAP data from an 8 min interval during the CME impact. The grey shaded regions indicate 8 magnetic flux ropes structures. }
    \label{fig:zoomzoom}
\end{figure}

The third and final feature we observe arising as the CME impacts are the large amplitude magnetic field spikes presented in Figure \ref{fig:zoom}. To investigate the nature of \textcolor{black}{these} spikes more carefully, we show in Figure \ref{fig:zoomzoom} a further zoomed in part of this interval, focusing on the early morning of 6 Oct 2015. In this figure, we now include the high-time resolution (57.8 Hz) ion current measured by LAP1, rather than the lower resolution sweep derived parameters. The contribution from photo-electrons has been \textcolor{black}{subtracted from this measured current, such that the measured current should be proportional to the ion flux. (The subtracted photo-electron current was about 25 nA and determined on a daily basis from the characteristics of the combined sweeps that day).}
The grey shaded regions in Figure \ref{fig:zoomzoom} indicate eight selected events, where the magnetic field strength increases to reach at least 150 nT in all but one case. The intervals are determined by eye as when the field is increased and the field orientation change occurs. The clock angle $= arctan(B_y/B_z)$ show clear rotations of the magnetic field in these intervals, while the cone angle $=arccos(B_x/|B|)$ shows an increase followed by a decrease during the events. The rotations are all in the same sense. The ion current typically increases a factor of about 1.5-3 during the spikes. However, the current increases are not always in concert with the magnetic field signatures. The current rather increases during a short interval within the field rotations periods, or at the edges of them. 

We have performed a minimum variance analysis (MVA) \citep{sonnerup1967} on these and several more similar spikes throughout the two day interval 5-6 Oct 2015. \textcolor{black}{The MVA is a single spacecraft method for obtaining the orientation of a stationary magnetic field structure in space, e.g. the normal of a current sheet or the axis of a magnetic flux rope. More specifically, the MVA gives the direction of minimum, maximum and intermediate variance of the magnetic field, which forms a right-handed coordinate system. The three orthogonal directions come from the eigenvectors of the co-variance matrix of the magnetic field components, calculated over a short time (the magnetic field spike, in this case). The eigenvector with the smallest associated eigenvalue is the minimum variance direction. The original magnetic field vectors can then be transformed into this new coordinate system so that, for example, one of the components is directed normal to the stationary structure assumed to exist in the space where the spacecraft is located.} 

\begin{figure}
	\includegraphics[width=\columnwidth]{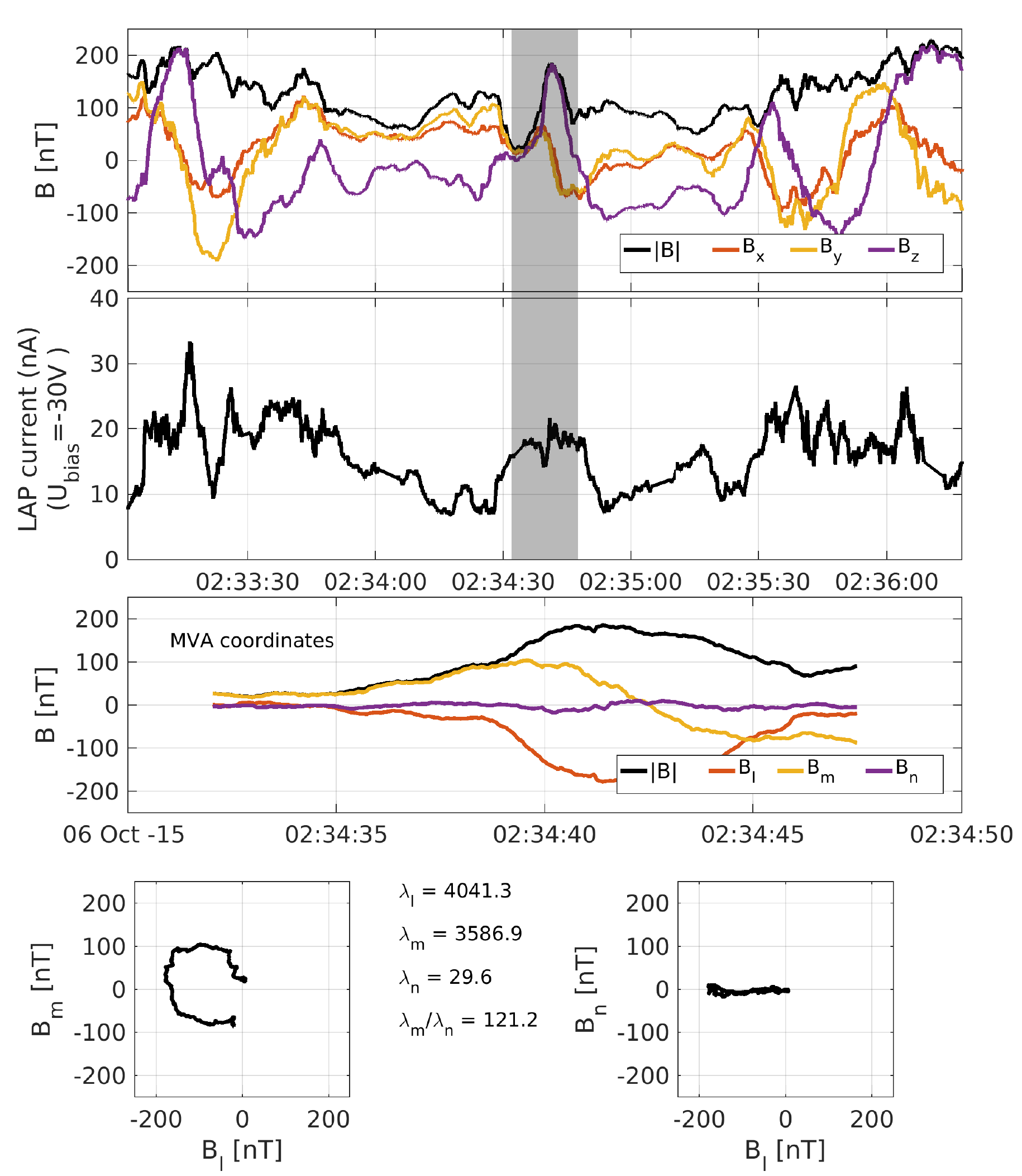}
    \caption{Results from an MVA analysis of one identified flux rope. The top panel shows the MAG data in CSEQ coordinates with the grey shaded region indicating the flux rope. The second panel shows the LAP1 ion current, which is proportional to ion density. The third panel shows the MVA coordinates from the time indicated by the grey shaded region, and the lower two panels show hodograms of the magnetic field components in MVA-coordinates. The eigenvalues as well as the ratio between the eigenvalues associated with the intermediate and the minimum components are also stated.}
    \label{fig:tworopes}
\end{figure}

\begin{figure}
	\includegraphics[width=\columnwidth]{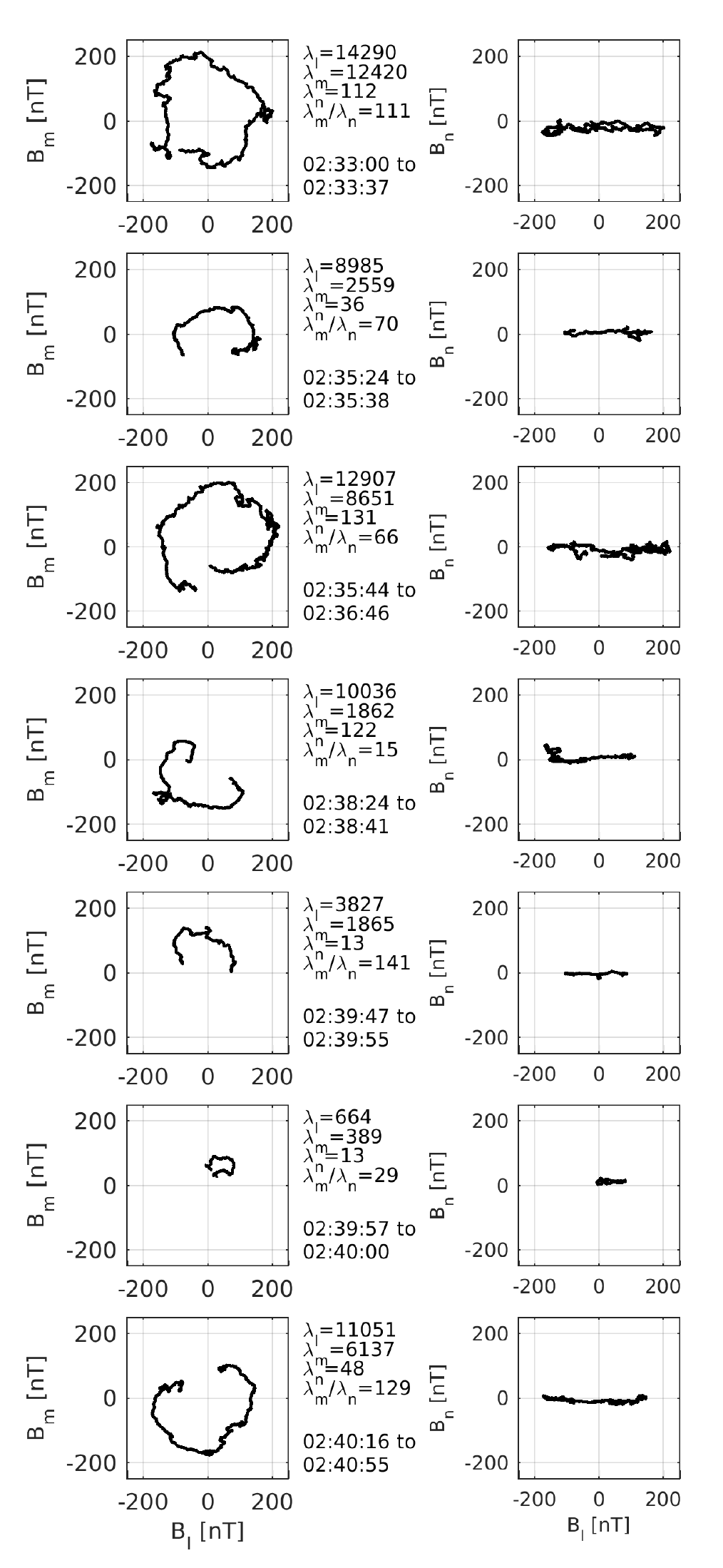}
    \caption{7 hodograms of the magnetic field data in MVA coordinates from the events indicated in Figure  \ref{fig:zoomzoom}. The eigenvalues as well as the ratio between the eigenvalues associated with the intermediate and the minimum components are written in the right panel for each event together with the time interval covered.}
    \label{fig:hodogram}
\end{figure}

From the results of the MVA, which we will show next, we find that at least 40 spikes appear to be magnetic flux ropes in this two day interval. Figure \ref{fig:tworopes} shows the results of the \textcolor{black}{MVA from one of the events presented in Figure \ref{fig:zoomzoom}. The grey shaded region again indicates the interval of the magnetic field spike/flux rope passage}. The hodograms of the magnetic field components in MVA-coordinates show characteristic circular pattern, if plotting the components in the direction of maximum and intermediate variance direction, and a near-straight line if plotting the maximum and minimum components, which are typical signature of flux ropes \citep[e.g][]{elphic1983}. Also, the eigenvalues $\lambda_l$, $\lambda_m$, and $\lambda_n$, associated with the maximum, intermediate and minimum eigenvectors (i.e. the direction of minimum, medium and maximum variance, respectively) are shown next to the hodograms. The ratio $\lambda_m/\lambda_n$ is well above 10, which is a limit for the accuracy of the determination of the eigenvectors.
Hodograms of the other seven events from Figure \ref{fig:zoomzoom} are shown in Figure \ref{fig:hodogram}, and they all have the similar characteristic shapes as the previous event.  


We note that not all spikes in the two day interval appear to be flux ropes, which indicates that the MVA analysis might not always work as intended and/or that there are also other dynamical processes at play, e.g. waves and other instabilities. We have not checked each individual spike throughout this interval, but rather focused on the largest amplitude spikes. We also note that there are current increases which are not associated with an identified flux rope but still with a field orientation change, e.g. at 02:35:20 UT and 02:37:30 UT in Figure \ref{fig:zoomzoom}. The identified flux ropes are not to be confused with the large-scale flux rope across the CME itself, but are rather short duration ($\sim$ 10-100 s) high amplitude flux ropes probably emanating from the solar wind interaction with the comet.

\textcolor{black}{Even though we have interpreted the large magnetic field spikes to be magnetic flux ropes, we do emphasize that there could be other possible explanations for these signatures, such as waves that have grown considerably in size.}

\begin{figure}
	\includegraphics[width=\columnwidth]{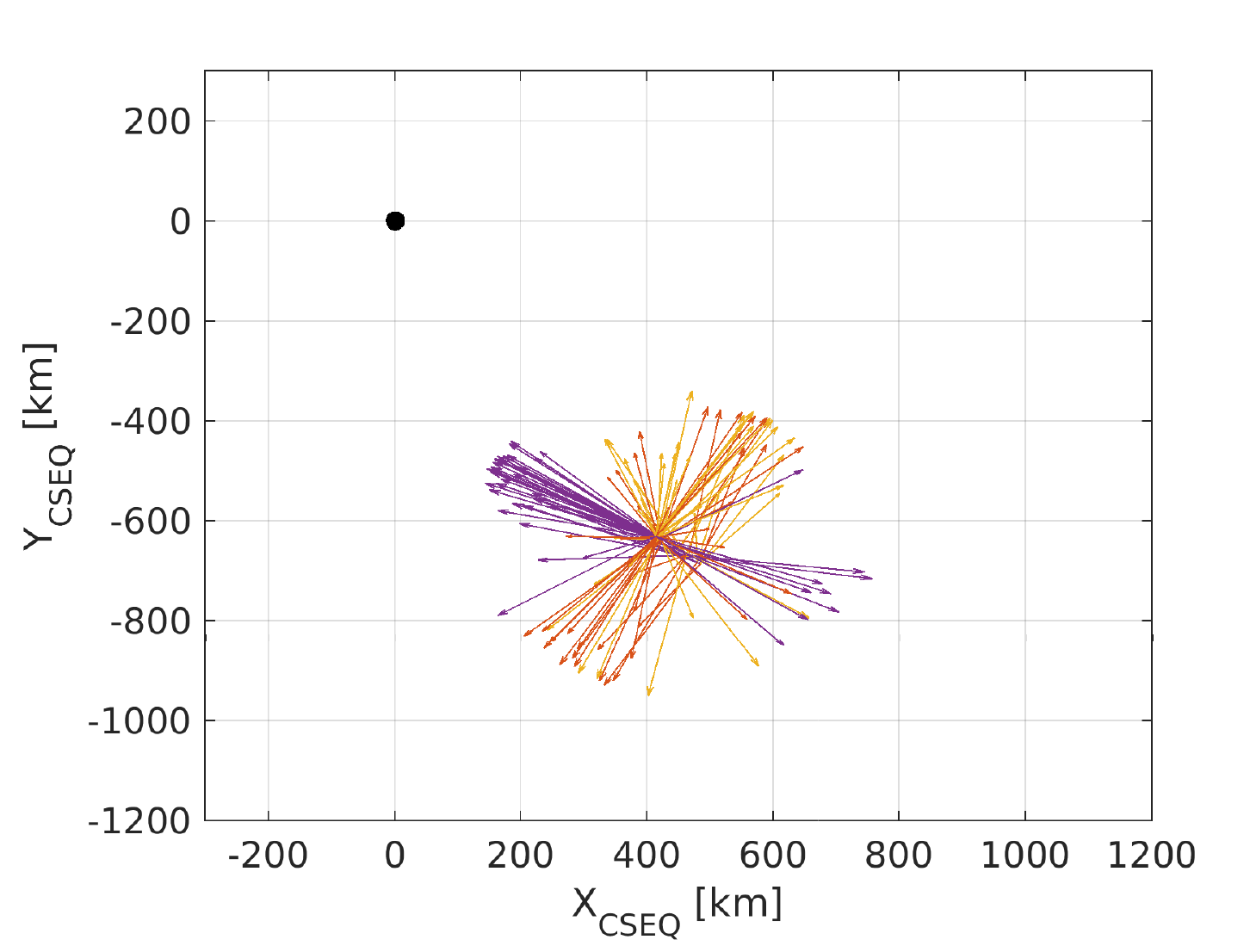}
    \caption{MVA vectors of 40 identified events in the interval 4 Oct 2015 - 6 Oct 2015. The purple vectors show the minimum variance direction, which are aligned with the rope axis. The red and the orange arrows indicate the maximum and intermediate variance direction, respectively, The 40 ropes all have very similar orientation and are directed slightly off the direction to the nucleus at (x,y)=(0,0)}
    \label{fig:orientation}
\end{figure}

\begin{figure}
	\includegraphics[width=\columnwidth]{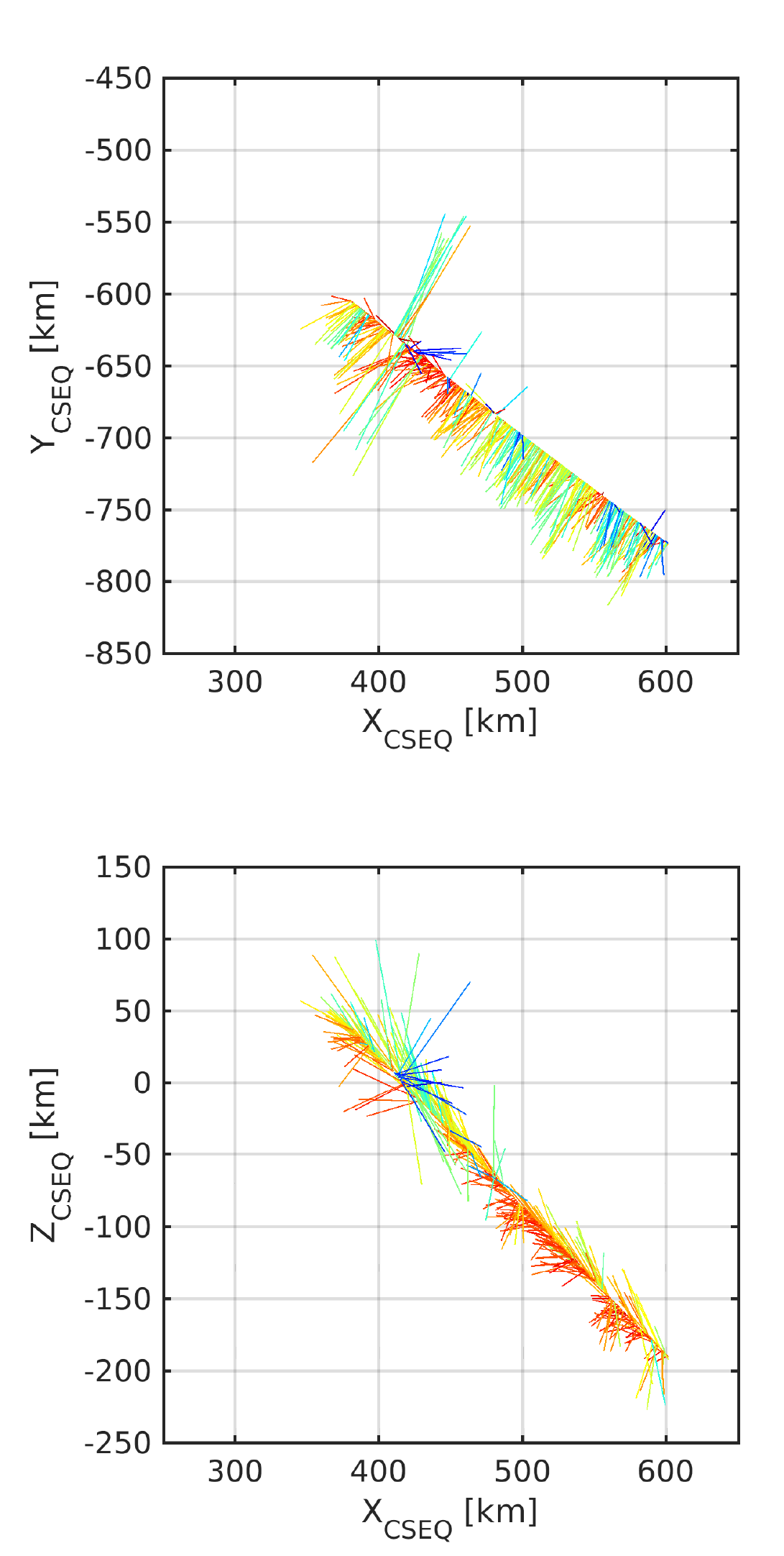}
    \caption{Magnetic field vectors projected on to the (top) x-y plane and (bottom) x-z plane along the Rosetta trajectory from 00:00 UT 4 Oct 2015 until 12:00 UT 6 Oct 2015. Red colours indicate an out-of-plane component, blue and in-to-plane component and green in-plane. The magnetic field is mainly directed in the -x,-y-direction before the large field rotations on 6 Oct 2015.}
    \label{fig:magvectors}
\end{figure}

For a flux rope, where the magnetic field is tightly wound up around its axis, the vector of the minimum variance direction gives the direction of the rope axis. Figure \ref{fig:orientation} shows the MVA vectors projected on the x-y plane for the selected 40 flux rope events. The axes along the rope, indicated by purple arrows, are quite ordered in their orientation and directed in between the comet nucleus and the anti-Sun direction.
The magnetic field direction in the interval around the time of the CME impact is shown in Figure \ref{fig:magvectors}, projected on the x-y and x-z planes. For most of the time, before the large event around midnight on 5 Oct, the field direction of the magnetic field is generally in the -x,-y-direction. Several orientation changes occur toward the end of this interval, when the CME main impact occurs, but before this the global field direction still has a preferred direction. The magnetic field direction is generally perpendicular to the axes of the identified flux ropes.

Some of the flux ropes observed here are unusually large for magnetic field magnitude and peak at over 200 nT in three events. These are peak field strengths larger than that of flux rope structures observed anywhere elsewhere in \textcolor{black}{interplanetary space}. However, larger flux ropes do exist in the solar corona. At Mars, large amplitude flux ropes have been observed to form through the interaction between crustal magnetic fields and the solar wind and reached peak magnetic field strength of 180 nT \citep{brain2010b}. However, the background field (outside of the flux rope) is at least 50 nT here, while for the event at Mars it was about 20 nT, so the relative increase is larger at Mars. In the ionosphere of Venus, flux ropes have been observed to reach about 100 nT \citep{zhang2012b}, although most are on the order of 10's of nT \citep{elphic1983}. The high amplitude flux ropes observed at the comet last typically for some 10's of seconds, although some can last as long as 100 s. Depending on the plasma bulk speed this will determine their physical sizes, if the flux ropes move with the plasma flow. The neutral gas outflow velocity is $\sim$700 m/s in the radial direction \citep{hassig2015}, and as the ions form primarily through ionisation of the neutrals their bulk flow will initially be roughly the same. However, it is not obvious that the flux ropes at the comet move with the neutral gas flow speed, since Rosetta was at this time at a distance of 800 km from the nucleus and in the ion-pile up region. In fact, IES measurements during the excursion indicate an ion velocity of $\sim$10 km/s, as stated in Section \ref{sec:density}. If the flux ropes are moving with speeds in the interval 1-10 km/s they would be on the order of 10-100's km large, in the direction of the flow (the velocity of Rosetta is insignificant). The flux ropes observed at Mars reported by \citet{brain2010b} and \citet{beharrell2012}, were on the order of $\sim$100 km, and the commonly observed flux ropes in the ionosphere of Venus  were found to have a radii of 6-15 km \citep{russell1979,elphic1983}. However, the high-amplitude ($\sim$40-100 nT) flux ropes observed at Venus are also on the order of 100 km \citep{zhang2012b}.

Furthermore, it is interesting to note that all the large amplitude flux ropes observed during the CME event have similar orientation, but still with some spread in direction. As noted before, Rosetta is practically standing still with respect to the comet while the plasma moves past it. The many flux ropes we observe, e.g. the large amplitude ropes in between 02:00-03:00 UT could then \textcolor{black}{potentially} be the same flux rope, which is moving back and forth past Rosetta. The magnetic field rotations mean that there is a current flowing in the flux rope which might make the rope kink and wobble as the current changes with distance. However, since the field rotations are always in the same sense this is probably not the case.

An interesting question is then how such flux ropes form. Closer to the nucleus, the diamagnetic cavity is observed intermittently. Sudden field topology changes should occur as the cavity forms and disappears, or moves radially, or as instabilities propagate along the boundary surface \citep{goetz2016}. Magnetic field spikes were also observed in close proximity to the magnetic cavity events, together with sudden density enhancements \citep{goetz2016, eriksson2016}. However, those were smaller in size and appeared to be more regular. It is possible that those are the same type of structures, but that the ones presented here have grown in size compared to the average cases close to the cavity, and been transported outward with the gas flow. During the transport outward they would likely become somewhat deformed. A possible scenario could be that as the CME impacts on the comet, the plasma environment is initially compressed and the density increase. As both the magnetic and thermal pressure consequently increases, and the CME pressure eventually decreases, the plasma regions and boundaries formed in the near-nucleus plasma could move outward. This would then lead to both the diamagnetic cavity growing and the flux ropes seen around the cavity events being transported outward. This is however somewhat speculative.

Flux ropes, which in principle are the same as magnetic islands, flux transfer events or plasmoids, typically form when there are large shears in a plasma, or, as an effect of magnetic reconnection. Large shears are certainly possible between the outflowing ionospheric plasma (1 km/s) and the solar wind flow (400 km/s). \citet{goetz2016} reported that the magnetic cavity events where probably associated with Kelvin-Helmholtz instabilities, during which flux ropes could also form. Magnetic reconnection is an other possible formation process. The ions are not magnetized in the coma while the electrons mostly are, which makes this an environment where magnetic reconnection processes could possibly occur. The largest amplitude flux ropes are observed in conjunction with the global field direction changes (see Figure \ref{fig:zoom}). A possible scenario would be that magnetic fields with different orientation meet as they convect through the coma, electrons decouple from the magnetic field, reconnection occurs and plasmoids/flux ropes are formed. Similar ideas have been proposed to occur at Venus when interplanetary magnetic field reversals propagate through the ionosphere of Venus \citep{edberg2011,vech2016}. The bursty nature of the magnetic field spikes during the CME impact, makes them appear similar to what has been observed in the magnetosphere of Mercury, where bursts of flux transfer events have been observed \citep{slavin2012}. Particle-in-cell simulations have shown that magnetic reconnection could form such bursts of plasmoids as the tearing instability disrupts the initial current sheet \citep{markidis2013}. However, further studies would be required to determine if magnetic reconnection could actually occur in the comet environment. 

Due to the orbit of Rosetta being on the dayside of the comet in this interval, we cannot study what is happening to the ion tail during the CME impact. Previous remote observations of comet-CME interactions have shown that tail disconnection events \textcolor{black}{can} occur as a CME impacts \citep{niedner1978,vourlidas2007,jia2009}. This process is usually attributed to magnetic reconnection in the comet tail. Here, we are possibly seeing  magnetic flux ropes being formed close to the nucleus instead. These are not the same as tail disconnection events although the processes involved could be similar. Some minor amount of plasma will still be carried away in the flux tubes. How much plasma is contained in the flux ropes is challenging to estimate since the exact scale and structure of them are somewhat unclear. However, if the density is on the order of 600 cm$^{-3}$ and the flux rope has a radius of about 100 km and is 600 km long, it would contain $\sim10^{20}$ particles. This assumes that each flux rope extends from 800 km down to 200 km (roughly where the magnetic cavity events were observed), and that the density is not decreasing with distance within the flux rope.

\section{Conclusions}
We have observed how a CME impacts on comet 67P when the comet was at 1.41 AU from the Sun (past perihelion). Rosetta was at this time on its inbound leg from the dayside excursion located at about 800 km from the nucleus. The plasma environment is significantly disturbed during the impact. The cold plasma density increases by as much as a factor of 10, to reach a maximum of 600 cm$^{-3}$, the suprathermal electron flux (10-200 eV) increases by a factor of 5-10, and the background magnetic field increases by a factor of \textcolor{black}{$\sim$2.5}, from about 40 to 100 nT, while individual magnetic spikes reach above 200 nT. 

The solar wind was observed to penetrate all the way down to 800 km during the CME impact. Previously, the solar wind was shielded from the deep coma since around April \textcolor{black}{2015}. When Rosetta was at 1500 km, and the solar wind conditions presumably normal, solar wind ions were not observed in Rosetta RPC/ICA or IES particle data. Hence, we conclude that the plasma environment was significantly compressed during the impact, due to the increased solar wind dynamic pressure. This is in agreement with the background magnetic field strength increasing by a factor of about 2-3, which would then explain the increased suprathermal electron fluxes as an effect of adiabatic compression \citep{madanian2016}.

The increase in cold plasma density is probably caused by a combination of compression of the global plasma environment, increased particle impact ionisation and charge exchange processes. As the CME impacts it is possible that Rosetta gains access to the cold and dense plasma located closer to the nucleus. The changing field topology across the CME might provide this possibility.

\begin{figure}
	\includegraphics[width=\columnwidth]{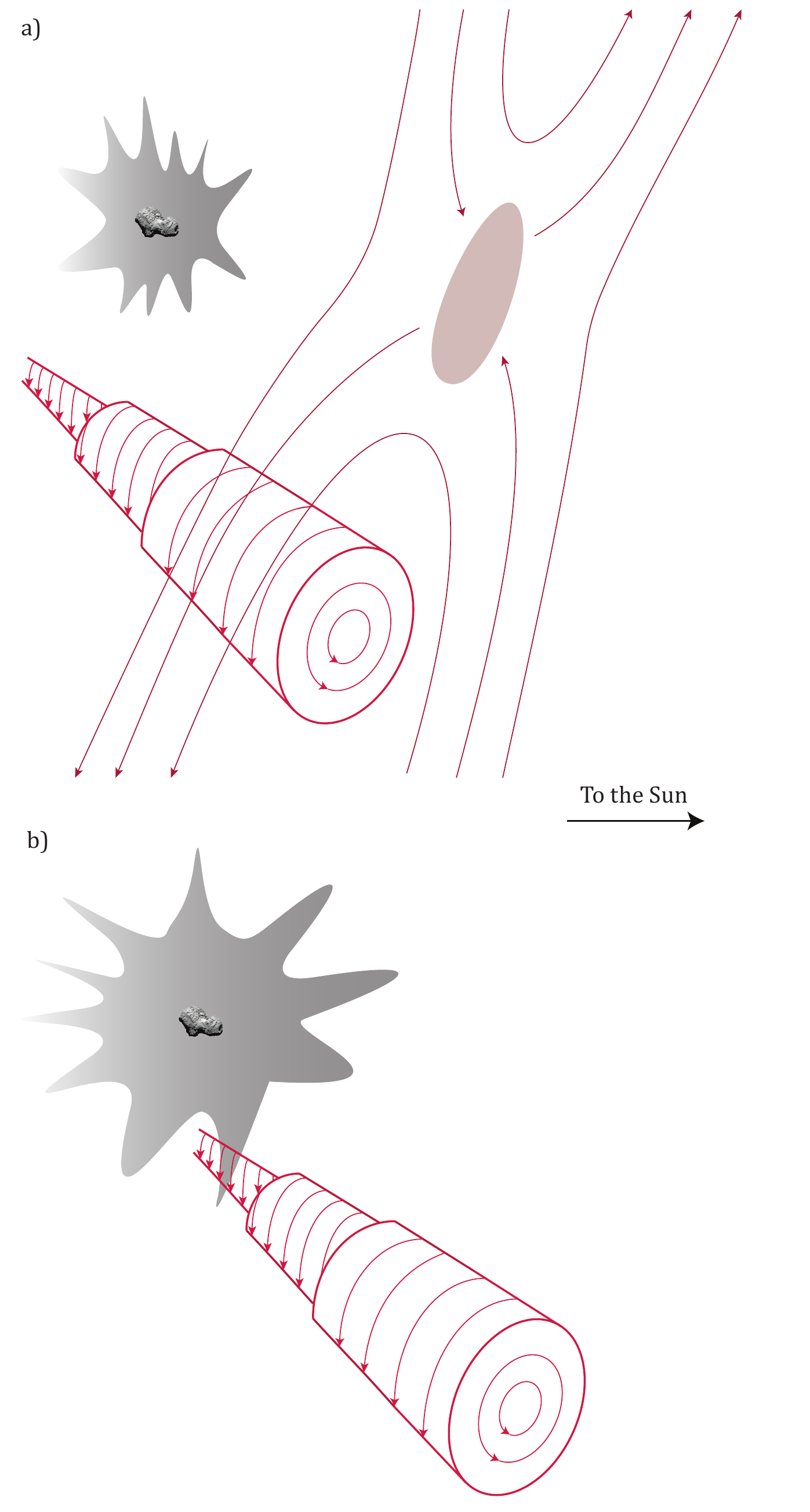}
    \caption{Schematic illustration showing how the observed magnetic field spikes could be interpreted as magnetic flux ropes. They could be associated with (a) magnetic reconnection initiated as the magnetic field reverses across the CME. When oppositely directed magnetic field configurations convect through the system and slow down, they eventually meet and reconnection can occur. Alternatively, the flux ropes are formed through (b) Kelvin-Helmholtz instabilities as a large shear arises in the plasma flow, which \citet{goetz2016} suggested to be associated with the diamagnetic cavity observations. The spiky nature of the near-nucleus coma is schematically depicted after the hybrid simulations results of \citet{koenders2015} (c.f. their Figure 6). The image of the comet nucleus is produced by the OSIRIS camera on Rosetta.}
    \label{fig:recon}
\end{figure}

Many of the magnetic field spikes are interpreted as magnetic flux ropes. An illustration aimed at explaining the formation mechanism of the flux ropes are shown in Figure \ref{fig:recon}. The flux ropes could be formed at this distance from the nucleus either through strong shears in the plasma, or as an effect of magnetic reconnection. Magnetic reconnection could occur when fields of different orientation pile up and meet in the coma as they convect through the system or, alternatively, the flux ropes form in close proximity to the diamagnetic cavity, where they appear more as 'spikey' waves or instabilities, such as Kelvin-Helmholtz instabilities, at first. They subsequently become significantly more extended as the CME impacts, but also become more twisted such that the field amplitude in the rope core increases to the extreme values.

\section*{Acknowledgements}
Rosetta is a European Space Agency (ESA) mission with contributions from its member states and the National Aeronautics and Space Administration (NASA). The work on RPC-LAP data was funded by the Swedish National Space Board under contracts 109/12, 135/13, 166/14 and 114/13 and Vetenskapsr\aa det under contracts 621-2013-4191 and 621-2014-5526. Support for RPC-MAG is provided by the German Ministerium f{\"u}r Wirtschaft und Energie and the Deutsches Zentrum f\"{u}r Luft- und Raumfahrt under contract 50QP 1401. The work on IES was supported in part by the U. S. National Aeronautics and Space Administration through contract 1345493 with the Jet Propulsion Laboratory, California Institute of Technology. Work at LPC2E/CNRS was supported by CNES and by ANR under the financial agreement ANR-15-CE31-0009-01. C.S.W. is supported by the Research Council of Norway grant No. 240000. This work has made use of the AMDA and RPC Quicklook database, provided through a collaboration between the Centre de Donn\'ees de la Physique des Plasmas (CDPP) (supported by CNRS, CNES, Observatoire de Paris and Universit\'e Paul Sabatier, Toulouse) and Imperial College London (supported by the UK Science and Technology Facilities Council). 




\bibliographystyle{mnras}
\bibliography{refs.bib} 








\bsp	
\label{lastpage}
\end{document}